\input psbox.tex
\splitfile{\jobname}
\let\autojoin=\relax
\documentstyle[preprint,revtex]{aps}

\def\BM#1{\mbox{\boldmath{$#1$}}}
\let\D=\Delta
\let\de=\delta

\let\Om=\Omega
\let\om=\omega
\let\ga=\gamma
\let\la=\lambda

\newcommand{\beps}{\bar\varepsilon}
\newcommand{\epsk}{\varepsilon_{\BM k}}
\newcommand{\bepsk}{\bar\varepsilon_{\BM k}}

\newcommand{\re}[1]{(\ref{#1})}
\newcommand{\non}{\nonumber \\}
\newcommand{\bib}{\bibitem}
\newcommand{\be}{\begin{equation}}
\newcommand{\ee}{\end{equation}}
\newcommand{\bea}{\begin{eqnarray}}
\newcommand{\eea}{\end{eqnarray}}
\newcommand{\ba}{\begin{array}}
\newcommand{\ea}{\end{array}}
\newcommand{\ra}{\rightarrow}

\input psbox.tex

\begin{document}

\begin{flushright}
{\large UBCTP CM1-1993\\May 1993}
\end{flushright}

\begin{title}
{\bf The Superfluidity and Experimental Properties \\ of Odd-energy-gap
Superconductors}
\end{title}

\vskip 15pt

\author{M. Dobroliubov$^a$, E.~Langmann$^b$,
P.C.E. Stamp}

\vskip 15pt

\begin{instit}
Physics Department, University of British Columbia \\ 6224 Agricultural
Road, Vancouver, BC, Canada, V6T 1Z1.
\end{instit}

\vskip 15pt

\begin{center} {\bf Abstract} \end{center}
\noindent
We consider the experimental properties of superconductors with a gap
which is an odd function of energy $\bepsk=\epsk - \mu$, i.e.\ , a gap
which vanishes everywhere on the Fermi surface; this is done within a
in a BCS framework. Apart from the standard phenomenology (density of
states, penetration depth, NMR, $C_V(T)$, $B_{c2}(T)$), we also look
at the stability of the superconducting state. It is stable to finite
supercurrents (although with a normal fluid density which increases
with the supercurrent density), and is also very weakly affected by
non-magnetic impurities. We find two classes of odd-gap superconductors,
which strongly differ in their low-energy properties. For a certain
parameter range, most of the results resemble those for d-wave
superconductors (except for the effect of impurities).

\vfill
\noindent
{\small $^a$ {Permanent address: Institute for Nuclear Research, Academy
of Sciences of Russia, Moscow 117312, Russia}\\
$^b$ {Erwin-Schr\"odinger fellow}
}

\newpage

It seems to have been first noticed some 30 years ago that for certain
models the BCS gap equation permits solutions in which the gap
function $\D_{\BM k}$ vanishes everywhere on the Fermi surface
\cite{Cohen}. Writing $\bepsk = \epsk - \mu$ (with $\epsk$ the fermion
band relation and $\mu$ the chemical potential) and ignoring all
frequency-dependence of the gap function \cite{Ber}, this means that
$\D(\hat{{\BM k}}, \bepsk=0) = 0$ for all directions $\hat{{\BM k}}$
onto the Fermi surface; the gap is then an odd function of $\bepsk$.
These solutions are then truly gapless, but nevertheless have a finite
stiffness to order parameter gradients, and, as we shall see, can
support a finite supercurrent density $\rho_s$ (although $\rho_s$ is
now a function of the superflow velocity $\BM v_s$).

Recently, odd-gap superconductors with a gap-functions behaving as
\be
\label{gap}
\D(\bepsk)=A(T)(|\bepsk|/\Om_c)^\ga{\rm
sign}(\bepsk) \mbox{ ~~~~for } |\bepsk|\ll\Om_c\quad (0\leq
\ga\leq 1)
\ee
with $\Om_c$ an upper cut-off, have been discussed in
the high-$T_c$ literature \cite{MilaAb} (as have odd-frequency gaps
\cite{Ber}) since they are favored by strongly repulsive short-range
interactions.  The corresponding density of states $N_s(\om)$ in the
superconducting state (at $T=0$) is depicted in Fig.1 and resembles
what is seen in tunneling experiments on HTSC (this was one of the
main results of
\cite{MilaAb}).

The aim of this paper is to elucidate other physical properties of
models leading to such states (\ref{gap}) (giving analytic results
where possible), and to give some understanding of their rather
remarkable stability under perturbations. To this end we shall not
only look at the standard list of experiments (specific heat, NMR,
ultrasound, critical field, penetration depth, etc.), but also
investigate the effect of impurities, and look at the dependence of
the superfluid density $\rho_s$ on the supercurrent density. As the
history of heavy fermion and high-$T_c$ experiments makes clear, it is
only by understanding the ensemble of experimental properties of a
given system that one can hope to pin down the pairing state
responsible; we hope the following results may facilitate the
examination of odd-gap candidates, as well as making their peculiar
physics somewhat clearer.

Note that we only consider odd-momentum gaps having $s$-wave symmetry
\cite{f1}. Moreover, for $\ga>1$, despite the presence of the Cooper
condensate in the ground state, the static current-current response
function vanishes at zero momentum transfer, so that there is no
Meissner effect and the system is not superconducting (superfluid).
Hence we restrict ourselves to models with $\ga\le 1$.

To allow analytic work, we choose a pairing interaction of
the following separable form
\be
V_{{\BM k},{\BM k}'}= V s(\beps_{\BM k}) s(\beps_{\BM k'})  \, ,
\ee
with
\be
s(\beps) = \left\{
\ba{cl}
\left(|\beps|/\Om_c \right)^\ga {\rm ~sign~} (\beps) \, ,&
{\rm ~if~} |\beps| <\Om_c
\cr
0 \, ,& {\rm ~if~} |\beps| \ge \Om_c  \, .
\ea
\right.
\label{s}
\ee
We believe that the essential properties of this model depend not on
the particular form of the potential (which we have chosen so as to
simplify the analytic work), but rather on the properties of the gap
produced by this potential. In other words, we expect that there
should exist a universality class for the momentum-odd gaps.  We shall
assume a cylindrical band-structure, $\bepsk = (k_x^2+k_y^2)/2m^*
-\mu$, and a constant density of states $N(0)$, modeling a layered
system. We shall also use the BCS gap equation, which for our model
results in $\D_{\BM k}=A(T)s(\bepsk)$ with $A(T)$ determined by
\bea
& \frac{1}{N(0)V} = \int_{-1}^{1} \frac{dy}{2e(y)}
\tanh\left(\frac{e(y)\Omega_c}{2 T}\right) |y|^{2\gamma} \, ,
\non
& e(y) = \sqrt{y^2+ \left(\frac{A(T)}{\Omega_c}\right)^2 |y|^{2\gamma}} \, .
\label{gap eq}
\eea
At the end of the paper we shall comment on the expected effect of
strong-coupling corrections. Note that several of our results for
$\ga\to 0$ are identical with the ones for the usual BCS model even
though these models are not the same.

At $T=0$, the gap equation has a non-trivial solution only if the
interaction is strong enough, i.e., if
\be
g_0\equiv N(0)V> 2\ga \, ;
\label{g0}
\ee
otherwise the system is normal down to $T=0$.

The dependence of the zero-temperature gap and the coupling strength
is given by (we use special functions as defined in \cite{AS})
\bea
&\frac{1}{g_0} = \frac{1}{(1+\gamma)}
\frac{1}{1+\sqrt{1+A(0)^2/\Omega_c^{2}}} \; \mbox{}_2F_1 \left( 1,
\frac{1}{1-\gamma}; \frac{2}{1-\gamma};
\frac{2}{1+\sqrt{1+A(0)^2/\Omega_c^{2}}} \right)
\non
\non
& = \left\{
\ba{cl}
\frac{1}{2\gamma} \left[ 1+\frac{\gamma}{1+\gamma} \frac{\pi}
{\sin (\frac{\pi}{1-\gamma})} \frac{\Gamma
(\frac{2}{1-\gamma})}{\Gamma^2 (\frac{1}{1-\gamma})}
\left( \frac{A(0)}{2\Omega_c} \right)^{\frac{2\gamma}{1-\gamma}}
+{\cal O}\left(\left(
{A(0)\over\Om_c}\right)^{\frac{4\gamma}{1-\gamma}}\right)
\right] ,
& {\rm~if~} \gamma < 1/2
\cr
\frac{1}{2\gamma} \left[ 1-\frac{\gamma}{2(2\gamma-1)}
\left( \frac{A(0)}{2\Omega_c} \right)^2  +
{\cal O}\left(\left( {A(0)\over\Om_c}\right)^4\right) \right] , &
{\rm~if~} \gamma > 1/2
\ea
\right.
\eea
The functional form of the corrections (${\cal O}\left(\left(
{A(T)\over\Om_c}\right)^{\frac{4\gamma}{1-\gamma}}\right)$ for
$\ga<1/2$, and ${\cal O}\left(\left(
{A(T)\over\Om_c}\right)^4\right)$ for $\ga>1/2$) is characteristic
for analytic expressions for our model, and for the rest of this paper
the symbol `$\simeq$' means equality up to such corrections.

The transition temperature is given by
\be
T_c = {(C[2\gamma])^{-1/2\gamma}\over 2} \Omega_c\left(1-
\frac{2\gamma}{g_0}\right)^{1/2\gamma} \, ,
\label{T$_c$}
\ee
with
 \be
C[x] = x 2^{1-x} (1-2^{1-x}) \Gamma (x) \zeta (x)
\ee
(note that for $\ga\ra 0$ this reduces to the standard BCS-$T_c$ eq.).
This function $C[x]$ is characteristic for this model and
enters many of the following explicit formulas.

The parameter $A(T)$, describing the magnitude of the gap function as
a function of temperature, is shown in Fig.2; it only changes slowly
with $\ga$. For $\ga\to 0$ one recovers the standard BCS result. For
$\ga=1$ we have
\be
A(T) \simeq U(g_0) \sqrt{1-t^2} \, ,~~~~t=T/T_c\, , ~~~~ \mbox{if }
U(g_0)\equiv 2\left( \left(\frac{g_0}{2}\right)^2 -1\right) \ll 1 \, .
\ee

We now discuss the main physical properties of this model, some of
which are rather surprising; most have not been previously described
(apart from the tunneling properties \cite{MilaAb}). The superfluid
density $\rho_s(T)/\rho_s(0)=\lambda_L^2(0)/\lambda_L^2(T)$
($\lambda_L(T)$ the London penetration length), calculated from the
linear current-current response in the usual way, is shown in Fig.3;
note that it behaves quite differently from $A(T)$, and varies quite
drastically with $\ga$!

It is worthwhile understanding this difference. Part of it lies in the
density of states $N_s(\om, T)$ in the superconducting state (cf.
Fig.1), and the related behavior of the coherence factor $\alpha(\om,
T)$ (in odd-gap superconductors, the coherence factors $\alpha_I$ and
$\alpha_{II}$ are equal!)
\be
 \alpha(\omega,T) =\int\; dE N_{s}(E)
N_{s}(E+\omega)\frac{1}{2\omega}
\left(f\left(\frac{E}{2T}\right) -
f\left(\frac{E+\omega}{2T}\right)\right) \, ,
\ee
(with $f(x)=1/({\rm e}^{x} +1)$) is shown in Fig.4. We notice that
$\alpha(T)$ only has a Hebel-Schlichter peak for the very restricted
range $\ga\ll 1$ (again, analytic results are easily obtained for
$\ga=1$, where $\rho_s(T)/\rho_s(0) \propto (1-t^2)$, and
$\alpha(T)\propto [1-U^2(g_0)(1-t^2)]$. However one also finds that
whereas $\rho_s(T=0)=\rho$ (the total superfluid density) for $\ga<1$,
for $\ga=1$, $\rho_s(T=0)=\rho[1-(1+A^2(0))^{-1}] <\rho$; we shall
return to this anomaly).

However another reason for this difference lies in the peculiar energy
scales in these odd-gap systems. In ordinary even-gap $s$-wave
superconductors it is well known that there is only one energy scale
that governs physical properties, i.e., $T_c$ (or what is the same
thing, $A(0)$). However this is not always true here; for $\ga>1/2$,
all physical properties depend on two energy scales which we can
choose to be $T_c$ and $\Om_c$, or using (\ref{g0},\ref{T$_c$}), to be
$T_c$ and $g_0$.  This is true even near $T_c$; for example the
superfluid density near $T_c$, usually measured using
$\rho_s(T)/\rho_s(0)=\la^2_L(0)/\la^2_L(T)$, is given analytically by
\bea
\frac{\lambda_L^2(0)}{\lambda_L^2(T)}
\simeq\left\{
\ba{cl}
 \frac{(2\gamma-1)^2}{\gamma(1-4\gamma)} \frac{C[2\gamma]
C[2\gamma-2]}{C[4\gamma-2]} \left[ \left(\frac{T_c}{T}\right)^2 -1
\right] , & {\rm~if~} \gamma < 1/2
\cr
\frac{(2\gamma-1)^2}{\gamma} C[2\gamma]C[2\gamma-2] \left( \frac{2
T_c}{\omega_c} \right) ^{2(2\gamma-1)} \left[
\left(\frac{T_c}{T}\right)^{2\gamma} -1 \right] ,
&{\rm~if~} \gamma > 1/2
\ea
 \right.
\label{London}
\eea
when $(T_c-T)\ll T_c$. Clearly the difference between these two
regimes is not obvious from numerical plots of physical quantities --
it is only revealed by the analytic expressions. Another way of
understanding it is to fix $T_c$ in equation \re{T$_c$}, and ask how
$g_0$ and $\Om_c$ then depend on each other; we see that $\ga=1/2$
marks the dividing line between weak and strong dependence, since
$dg_0^{-1} \propto (T_c/\Om_c)^{2\ga-1} d\Om_c/\Om_c$ (recall that in
the weak coupling case $T_c\ll\Om_c$).

We can also evaluate other response functions of the system; the most
interesting are the specific heat $C_v(T)$ (Fig.5), and the upper
critical field $B_{c2}(T)$ for a magnetic field parallel to the
cylinder ($z$-) axis (Fig.6). The size of the specific heat jump has
the analytic form
\be
 \frac{\Delta C_V(T_c)}{T_c}
 \simeq \left\{
\ba{cl}
 \frac{4(1-2\gamma)}{4\gamma-1} \frac{C[2\gamma]^2}{C[4\gamma-2]}
 {\rm~if~} \gamma < 1/2
\cr
4(2\gamma-1) C[2\gamma]^2 \left( \frac{2 T_c}{\Omega_c} \right)
^{2(2\gamma-1)} &{\rm~if~} \gamma > 1/2
\ea
\right.
\ee
which again shows the two relevant energy scales for $\ga>1/2$.

The reduced upper critical field
\be
b^*_{c2}(t) = \left.\frac{B_{c2}(tT_c)}{-T_c{\rm d}B_{c2}(T)/{\rm d}
T}\right|_{T=T_c} \quad (0\leq t\leq 1)
\ee
is determined by
\be
\frac{1}{2\gamma}\left(\left(\frac{1}{t}\right)^{2\gamma}-1\right) =
\Phi_\gamma(b_{c2}^*(t)/t^2)
\ee
with
\be
\Phi_\gamma(\xi) =  \frac{1}{C[2\ga]}\int d\! y y^{2\gamma}
\frac{\tanh(y)}{2y}\int\frac{d\! x}{\sqrt{2\pi}} e^{-x^2/2}
 \frac{\sinh^2(x\sqrt{c_{\gamma} \xi})}
{\sinh^2(x\sqrt{c_{\gamma} \xi}) + \cosh^2(y)} \, ,
\ee
and the constant $c_{\gamma} = (2\ga-1) C[2\ga-2]/2C[2\ga]$.
Moreover, the slope is
\be
T_c\frac{{\rm d}B_{c2}(T)}{{\rm d}T}|_{T=T_c} = -16 c_\ga
\frac{T_c^2}{e v_F^2}
\ee
with $v_F=\sqrt{2\mu/m^*}$ and $e$ the elementary charge.  For $\ga\to
0$ one recovers the BCS result for a cylindrical band structure
\cite{EL}. The full $b_{c2}^*(t)$-curves for different $\ga$-values
are given for completeness in Fig.6. We note that one should trust
this result only for not too large magnetic fields --- otherwise the
phase approximation used in the calculation is poorly justified
\cite{EL}.  For HTSC this amounts to restricting to temperatures close
to $T_c$.  However, this is also the only regime where reliable and
accurate experimental data are accessible.

This summarizes our investigation of the common response functions for
these superconductors. However a crucial question is still unanswered,
viz., how can we be sure that the superfluid state is stable? Despite
the demonstration in linear response theory (i.e., for infinitesimal
currents) of a stiffness to transverse gauge perturbations (Fig.3),
the complete gaplessness of these states makes it necessary to look at
how superflow breaks down at finite current density; one must also
show that ordinary non-magnetic impurities do not have a catastrophic
effect on superflow.

Given a finite superflow velocity ${\BM v}_s$, the quasiparticle
spectrum in the laboratory (or sample) frame will be $E_{\BM p}({\BM
v}_s) = E_{\BM p} -{\BM p}{\BM v}_s$, where the quasiparticle spectrum
is $E_{\BM p}^2 = \beps^2_{\BM p} +|\D_{{\BM k}}|^2$ as usual. Just as in
$^3$He-A, the superfluid resolves the problem of a zero Landau
critical velocity by exciting quasiparticles until the new Galileo
transformed Fermi surface (shifted by wave-vector ${\BM q} = m^* {\BM
v}_s/\hbar$ from the lab frame) has been filled by them. At this point
the exclusion principle prevents any further quasiparticle excitation.
Thus we find that even at $T=0$, a finite supercurrent density will
give rise to a finite normal fluid density. The calculation of
$\rho_N(T,{\BM v}_s)$ requires numerical work, but one is most
interested in $\rho_N(T=0,{\BM v}_s)$ for small ${\BM v}_s$; one then
easily finds that $\rho_N({\BM v}_s)/\rho \propto \left( p_F
v_s/A(0)\right)^{{1\over\ga}-1}$. Thus, the superfluid is stable to
finite supercurrent excitations.

Another possible source of zero temperature normal fluid is impurity
scattering; this is well known to lead to zero energy resonant states
in higher-$\ell$ paired states \cite{imp}. However we now demonstrate
that the effect of non-magnetic scatterers on these states will be
very small. This is conveniently done by examining the impurity
T-matrix in the dilute limit; as a function of complex frequency $z$
one has \cite{imp}
\be
\hat{T}_S ({\BM k},{\BM k}^{\prime}; z) =
\hat{T}_N ({\BM k},{\BM k}^{\prime}; z) +
\sum_{{\BM k}^{\prime\prime}} \hat{T}_N ({\BM k},{\BM k}^{\prime\prime}; z)
\left[ \hat{G}_S({\BM k}^{\prime\prime},z)-\hat{G}_N({\BM k}^{\prime\prime},z)
\right] \hat{T}_N ({\BM k}^{\prime\prime},{\BM k}^{\prime}; z)
\label{T-matrix}
\ee
where $S,N$ label the superconducting and normal functions, and refers
to the 1-particle Green function matrix in the {\em pure state}; we
work with Nambu notations. The basic physics is revealed using a
simple $s$-wave impurity scatterer, with normal state phase shift
$\de_0=0$; then $\hat{T}_N ({\BM k},{\BM k}^{\prime}; z)\rightarrow
\hat{T}_N (z)$, which is diagonal with elements
\be
t_N(z) = -\frac{1}{\pi N(0)} e^{i\de_0} \sin{(\de_0)} \, .
\ee
 From \re{T-matrix} we then have the solution
\bea
\hat{T}_S(z)=\hat{T}_N(z) \left[ 1-\hat{M}(z)\hat{T}_N(z)\right]^{-1} \, ,
\non
\hat{M}(z) = \sum_{{\BM k}} \left[ \hat{G}_S({\BM k},z)-\hat{G}_N({\BM k},z)
\right] \, .
\eea
Carrying out the integration, and making the continuation
$z\rightarrow\om+i\de$, one finds the diagonal $\hat{T}_S$ with
elements
\be
t_S(\om+i\de) = \frac{-i}{\pi N(0)}
\frac{(1-\ga)\Om^2_0(\om)+\ga\om^2}{
(1-\ga)\Om^2_0(\om)+\ga\om^2-|\om|\Om_0(\om)
\frac{\sin{(\de_0)}}{1+\sin{(\de_0)}}}
\frac{\sin{(\de_0)}}{(1+\sin{(\de_0)})} \, ,
\ee
where $\Om_0(\om)$ is the root of the equation
\be
(1-\ga)\Om^2_0(\om)+A^2(T)\Om_0^{2\ga}(\om) = \om^2 \, .
\ee
 Now this $T$-matrix has an impurity state pole only for $\ga\le1/2$,
at a energy which is never zero, even in the unitarity limit
$\de_0=\pi/2$ ( for $\de_0\ll 1$, $\Om_0$ lies near the maximum of
$N_s(\om)$, which also exist only for $\ga\le 1/2$). From this we
conclude that no low energy states are induced by scattering in the
dilute limit. The effect of a finite impurity concentration can be
modeled using the CPA extension of the above method \cite{CPA}, and
leads to an additional contribution to $N_S(\om)$, smeared out around
$\om=\Om_0$. Just as in conventional $s$-wave even gap systems
\cite{And}, non-magnetic scatterers are thus seen to have a benign
effect on superconductivity.

We may now summarize the general properties of these odd-momentum gap
states as follows. They divide into two distinct classes, which we
might call "infra-red weak" (for $0\le\ga\le 1/2$), and "infra-red
strong" (for $1/2\le\ga\le 1$). The IR weak case has a density of
states which disappears faster than linear as $\om\ra 0$, and has an
obvious gap structure in $N_S(\om)$; the low-$\om$ properties are
governed only by the energy scale $T_c$, as in conventional even-gap
$s$-wave superconductors (and indeed as $\ga\ra 0$, we end up with the
same $N_S(\om)$). The IR strong systems, in contrast, are governed by
two energy scales, and have a density of states with no obvious gap
structure, disappearing more slowly than linear. Some experimental
properties vary only slowly with $\ga$ (such as $A(T)$, or
$B_{c2}(T)$), and so do not look radically different from conventional
superconductors; other properties are very strongly dependent on $\ga$
(such as $C_v(T)$, $\D C_v(T)$, ultrasonic attenuation, NMR,
$N_S(\om;T)$, and $\la^2_L(T)$), and look more and more different from
conventional superconductors as $\ga$ increases from zero; for
$\ga>1/2$ they look completely different. Superflow is stable, even
with impurity scattering, but only at the cost of a
supercurrent-dependent normal fluid density $\rho_N (v_s,T)$, which is
finite even at $T=0$, if $v_s$ is finite.

Readers familiar with the experimental properties of high-$T_c$
superconductors will immediately recognize the similarity between the
plots in Figs.1 and 3--5 and the data for, eg., the Y-based compounds
\cite{exp}; this resemblance is particularly striking for $\ga\sim
0.28$. In fact for values of $\ga$ around this value, one finds a
striking resemblance to the properties of "$d$-wave superconductors"
(ie., single even-gapped states with line nodes). Both states then
have $\rho_s(T)\propto T$ at low $T$, no Hebel-Schlichter peak in the
NMR (and with a roughly linear dependence on $T$ except for $t\ll 1$),
a density of states $N_S(\om)\propto \om$ at low $\om$, and a somewhat
reduced specific heat jump at $T_c$ (with non-exponential behavior for
$t\ll 1$). Their $B_{c2}(T)$ curves are also similar; where they
differ is the greatly increased sensitivity of $d$-wave states to
non-magnetic impurity scattering (and in fact this is the principal
stumbling block, at present, to a $d$-wave theory; the high-$T_c$
superconductors are often remarkably insensitive to such scattering).
Another difference will arise in their tunneling, through junctions
into conventional superconductors \cite{d-wave}

It thus seems at least reasonable to pursue explanations of high-$T_c$
superconductivity in terms of these odd-gap states \cite{f2}. However
we would like to finish with a cautionary note. We have nowhere
incorporated strong-coupling into our calculations, and it is not
entirely obvious what these would do. One might speculate, on the
basis of experience with Fermi-liquid corrections \cite{Leg} for
non-conventional pairing states, that the general effect would be to
narrow features such as the specific heat peak, or the rise in
$\rho_N(T)$, near T$_c$ (this is what happens in $^3$He). However an
Eliashberg-type analysis may give a different answer \cite{Carb}. In
fact it turns out that employing an Eliashberg analysis is
non-trivial, because it leads to a coupled system of equations:
Besides the usual BCS equation \re{gap eq} with a {\em renormalized}
band-relation $\tilde{\varepsilon}_{{\BM k}}=\bar\varepsilon_{{\BM
k}}-R_{{\BM k}}$ one gets an equation for the band renormalization,
\be
R_{{\BM k}}=\sum_{{\BM k'}}V_{{\BM k}{\BM k'}}n_{{\BM k}'},
\ee
with $n_{{\BM k}}=\frac{1}{2}(1-\tilde\epsk\tanh(E_{\BM k}/2T)/E_{\BM
k})$ ($E_{{\BM k}}=\sqrt{\tilde\epsk^2+|\D_{{\BM k}}|^2}$) the
distribution function in the superconducting state. For effective
potentials $V_{{\BM k}{\BM k}'}$ which are essentially constant in the
vicinity of the Fermi surface, $R_{{\BM k}}$ is irrelevant as it can
be absorbed into the chemical potential. This is not the case for our
separable potential.  One gets $R_{{\BM k}}= R(T) s(\varepsilon_{{\BM
k}})$ which apparently is significant in the infra red.  Though we
think that our simplified analysis takes into account the most
important qualitative features of energy dependent interactions, it
would be interesting to solve the coupled system of equations
determining the parameters $A(T)$ and $R(T)$.

{\bf Acknowledgements} M.D. and P.C.E.S. were supported in part by the
Natural Sciences and Engineering Research Council of Canada.  E.L.
was supported by the ``Fonds zur F\"orderung der wissenschaftlichen
Forschung'' under contract No. J0789-PHY. We thank
J.C.Carbotte, W.Hardy, N.Prokof'ev and especially D.Bonn for discussions.

\newpage

\begin{center}
{\large\bf Figures}
\end{center}

\noindent
 The curves on the figures correspond to the following values of $\ga$:
\begin{center}
 $\gamma=0$ -- solid line, $\gamma=1/4$ -- short-long dashed line,
 $\gamma=1/2$ -- long-dashed line, \\ $\ga=3/4$ -- short-dashed line,
$\gamma=1$ -- dotted line.\\
\end{center}
We assumed that $T_c/\Om_c = 0.1$

\begin{center}

$$\psboxto(4.5in;3.5in){ns.eps}$$
{\small {\bf Fig.1}. The profiles of the zero-temperature density of states
$N_s(\om)$.}

$$\psboxto(4.5in;3.5in){a.eps}$$
 {\small {\bf Fig.2}.
The temperature dependence of the gap parameter $A(T)$.}

$$\psboxto(4.5in;3.5in){lon.eps}$$
{\small {\bf Fig.3}.
The temperature dependence of the London penetration depth
$\la_L^2(0)/\la_L^2(T)$.}

$$\psboxto(4.5in;3.5in){alpha.eps}$$
{\small {\bf Fig.4}.
The temperature dependence of the coherence factor $\alpha(T)$.}

$$\psboxto(4.5in;3.5in){cv.eps}$$
{\small {\bf Fig.5}.
The temperature dependence of the specific heat $C_V(T)/T$.}

$$\psboxto(4.5in;3.5in){bc.eps}$$
{\small {\bf Fig.6}.
The temperature dependence of the reduced magnetic field $b^*_{c2}$.}

\end{center}

\autojoin

\end{document}